\documentclass{aa}
\usepackage{graphicx,natbib}
\newcommand{\an}{{Astr.~Nachr.}}			
\newcommand{\dfrac}[2]{\displaystyle\frac{#1}{#2}}

\newcommand{\scri}{\scriptsize}

\newcommand{\rbcz}{r_{\mbox{\scriptsize bcz}}}
\newcommand{\rin}{r_{\mbox{\scriptsize in}}}
\newcommand{\rmc}{r_{\mbox{\scriptsize mc}}}
\newcommand{\Bconf}{B_{\mbox{\scriptsize conf}}}
\newcommand{\Pcyc}{P_{\mbox{\scriptsize cyc}}}
\newcommand{\Omegabcz}{\Omega_{{\mbox{\scriptsize bcz}}}}

\newcommand{\Prm}{\mbox{Pr}_{{\mbox{\scriptsize m}}}}
\newcommand{\ov}{\overline}
\newcommand{\omegacyc}{\omega_{\mbox{\scriptsize cyc}}}
\newcommand{\Hskin}{H_{\mbox{\scriptsize skin}}}

\begin{document}
   \title{Dynamics of the fast solar tachocline}

   \subtitle{I. Dipolar field}

   \author{E. Forg\'acs-Dajka\inst{1}
          \and
          K. Petrovay\inst{1,2}
          }

   \offprints{E. Forg\'acs-Dajka}

   \institute{\inst{1}E\"otv\"os University, Department~of Astronomy, Budapest, Pf.~32, 
              H-1518 Hungary\\
	      \inst{2}Instute for Theoretical Physics, University of California,
	      Santa Barbara, CA 93106-4030, USA\\
              \email{E.Forgacs-Dajka@astro.elte.hu, K.Petrovay@astro.elte.hu}
             }

   %\date{Received 15 January 2002 / Accepted 16 April 2002}
   \date{Astronomy and Astrophysics, v.389, p.629-640 (2002)}

\abstract{
One possible scenario for the origin of the solar tachocline,  known as the
``fast tachocline'', assumes that the turbulent diffusivity exceeds $\eta\ga
10^9\,$cm$^2/$s. In this case the dynamics will be governed by the
dynamo-generated oscillatory magnetic field on relatively short timescales.
Here, for the first time, we present detailed numerical models for the fast
solar tachocline with all components of the magnetic field calculated
explicitly, assuming axial symmetry and a constant turbulent diffusivity $\eta$
and viscosity $\nu$. We find that a sufficiently strong oscillatory poloidal
field with dipolar latitude dependence at the tachocline--convective zone
boundary is able to confine the tachocline. Exploring the three-dimensional
parameter space defined by the viscosity in the range $\log\nu=9$--$11$, the
magnetic Prandtl number in the range $\Prm=0.1$--$10$, and the meridional flow
amplitude ($-3$ to $+3\,$cm$/$s), we also find that the {\it confining field
strength} $\Bconf$, necessary to reproduce the observed thickness of the
tachocline, increases with viscosity $\nu$, with magnetic Prandtl number
$\nu/\eta$, and with equatorward meridional flow speed. Nevertheless, the
resulting $\Bconf$ values remain quite reasonable, in the range
$10^3$--$10^4\,$G, for all parameter combinations considered here. The
thickness of the tachocline shows a marked dependence on both time and
latitude. The latitude dependence is similar to that inferred by
helioseismology, while the time dependence is within the observational errors.
\keywords{Sun: interior, MHD, Sun: rotation}
}

   \maketitle
%
%________________________________________________________________

\section{Introduction}

The tachocline is the thin transitional layer below the solar convective zone
between the surface-like differential rotation pervading the convective zone
and the rigid rotation of the radiative interior. The existence and properties
of this layer have been known from helioseismic studies, recently reviewed by
\citet{Corbard+:SOGO}. The tachocline is known to be
extremely thin \citep{Corbard+:AA98,Corbard+:AA99,Char+:tacho.thickness}. 
Precise values for its thickness depend on the particular fitting profiles used
in helioseismic forward modelling for the residual rotation rate
$\Omega-\Omega_0$ (where $\Omega_0$ is the rotation rate of the solar
interior). The scale height of $\Omega-\Omega_0$ is generally found to be
crudely $0.01$--$0.02\,R_\odot$, so that the residual rotation is reduced by
one order of magnitude within a layer of barely $\sim 30\,$Mm. Near the equator
this layer lies just below the adiabatically stratified convective zone. 
Inversions seem to suggest that at 
higher latitudes it is situated at a slightly but significantly higher level,
and is apparently also thicker, so that it partly overlaps with the convective
zone \citep{Basu+Antia:tachovar}.

The extreme thinness of the tachocline implies a strongly anisotropic
(predominantly horizontal) transport of angular momentum. Several different
mechanisms have been proposed for this transport, but it is now widely believed
that the magnetic field is instrumental in its origin. Magnetic fields can lead
to the required anisotropic transport either directly, via the action of
Maxwell stresses, or indirectly, by rendering the rotation
profile unstable, and thus giving rise to anisotropic turbulent transport 
\citep{Gilman+Dikpati:unstable.tacho,Gilman:tacho.review}. (In 
contrast, purely hydrodynamic mechanisms are apparently unable to provide the 
necessary lateral transport --- cf. \citet{Char+McGregor:tacho.HDstab,
Garaud:tacho.instab}.)

Depending on the value of the magnetic diffusivity, this magnetic field may
either be a weak permanent, primordial field pervading the solar interior, or
the strong oscillatory field generated by the solar dynamo. A magnetic field
oscillating with a circular frequency $\omegacyc=2 \pi /P$,  $P= 22$ years is
known to penetrate a conductive medium only down to a skin depth of 
\[ \Hskin=(2\eta/\omegacyc)^{1/2} \label {eq:skin} \]  
where $\eta$ is the magnetic diffusivity. 
Comparing this with the tachocline scale height quoted above, it follows that 
for $\eta\la 10^8\,$cm$^2/$s the dynamo field cannot penetrate the tachocline,
and we can expect the tachocline to be pervaded by the internal primordial
field. On the other hand, for $\eta\ga 10^9\,$cm$^2/$s the tachocline dynamics
should be governed by the dynamo field. As the associated diffusive and Lorentz
timescales are also very different, these two cases basically correspond to the
case of ``slow'' and ``fast'' tachocline, discussed in the literature 
\citep[see esp. Table I in][]{Gilman:tacho.review,Brun:SOGO}.

The case of a slow tachocline \citep{Gough+McIntyre} has been studied
extensively in recent years by a number of authors. In models with no
meridional flow \citep{McGregor+Char, Rudiger+Kichat:thin.tacho} it was found 
that for purely molecular diffusivities even a rather weak prescribed internal
poloidal field of $10^{-3}\,$G is sufficient to confine the tachocline to its
observed thickness if the internal field is fully contained within the
radiative zone. The steady toroidal field resulting from the winding up of
poloidal field lines may reach kilogauss values. The containment of the
internal field is, however, not so easy to explain, given that the surrounding
convective zone, with its high diffusivity, is far from being a perfect
conductor. On the other hand, for an  internal field not contained within the
radiative zone Ferraro's law of isorotation predicts that the differential
rotation will penetrate deep into the radiative zone. This was indeed confirmed
in the numerical models. Recently, \citet{Garaud:tacho1} performed calculations
taking into account the meridional flow and the self-consistent evolution of
the poloidal field (with otherwise simplified physics). Her models show that
there exists an intermediate range of field strengths where the interaction
with the meridional flow indeed leads to the poloidal field being mostly
confined to the radiative zone, with a resulting rotation profile that is
comparable to the observations. At any rate, these calculations stress the
importance of including meridional flows and a self-consistent treatment of the
poloidal field.

The alternative case of a fast tachocline, in contrast, has received much less
attention. This is so despite that several authors,
\citep[e.g.][]{Gilman:tacho.review} noted that the tachocline is more likely to
be dynamically coupled to the convective zone than to the radiative zone, and
that a part of the tachocline is likely to overlap with the strongly
turbulent convective zone, especially at higher latitudes. Faced with this
situation, in a recent paper (\citet{Dajka+Petrovay:mgconf}, hereafter FDP01)
we considered the case of a turbulent tachocline with a turbulent diffusivity
of about $\eta=10^{10}\,$cm$^2$/s, pervaded by an oscillatory dipole field. This
model can be regarded as the analogue of the models of
\citet{Rudiger+Kichat:thin.tacho} and \citet{McGregor+Char} for a fast
tachocline. It was found that for a sufficiently high value of the field
strength, Maxwell stresses in the oscillatory field transported angular
momentum efficiently enough to confine the tachocline to its observed
thickness. The poloidal field strength necessary for this confinement could
also be estimated from a simple analytic relation; with the diffusivity quoted
above its value was $2400\,$G.

The analysis of FDP01 was of limited scope, as several simplifying assumptions
were made. Only the case of magnetic Prandtl number $\Prm=\nu/\eta=1$ ($\nu$ is
the viscosity) was considered, meridional circulation was neglected, the
poloidal field was prescribed throughout the computational domain, the applied
field was an oscillating dipole instead of a dynamo wave, the diffusivities
were arbitrarily prescribed as constant or known functions of radius etc. The
aim of the present series of papers is to systematically generalize the model
of the fast tachocline by incorporating more and more of the effects neglected
in FDP01. As a first step, in the present paper we will incorporate the
meridional flow and include a self-consistent calculation of the poloidal
field, the importance of which was underlined by the results of
\citet{Garaud:tacho1} in the context of the slow tachocline. We also explore
the parameter space a bit more widely by varying the magnetic Prandtl number
and the diffusivity. 

After formulating the mathematical problem in Sect.\,2, results of the model
calculations will be presented in Sect.\,3. Finally, the Conclusion summarizes
the main results.

%__________________________________________________________________

\section{The model}
\subsection{Equations}

The time evolution of the velocity  field $\mathbf{v}$ and magnetic field
$\mathbf{B}$ are governed by the Navier-Stokes and induction equations. We
write these in a frame rotating with the fixed internal rotation rate
$\Omega_0$:
\begin{eqnarray}
\label{eq:NS}
\partial_t \mathbf{v}&&+ \left( \mathbf{v} \cdot \nabla \right) \mathbf{v} = 
         \nabla \left(\frac 12|\vec\Omega_0\times\mathbf{r}|^2 -V\right) 
	 -2(\vec\Omega_0\times\mathbf{v}) \nonumber \\ 
  &&- \dfrac{1}{\rho} \nabla \left( p+ \dfrac{B^2}{8\pi} \right) 
+ \dfrac{1}{4\pi\rho}(\mathbf{B} \cdot \nabla)\mathbf{B} + \dfrac{1}{\rho}
\nabla\cdot \hat{\tau} ,           
\end{eqnarray}
and
\begin{equation}
\partial_t \mathbf{B} = \nabla \times ( \mathbf{v} \times \mathbf{B} ) - 
  \nabla \times ( \eta
  \nabla \times \mathbf{B}) ,               \label{eq:induc}
\end{equation}
where $\mathbf{r}$ is the radius vector, $\hat\tau$ is the viscous stress
tensor, $\eta$ is the (turbulent) magnetic diffusivity, $V$ is the
gravitational potential and $p$ is pressure. These equations are supplemented
by the constraints of mass and magnetic flux conservation
\begin{equation}
\nabla\cdot ( \rho \mathbf{v}) = 0 ,
\end{equation}
\begin{equation}
\nabla \cdot\mathbf{B} = 0 .
\end{equation}

We assume spherical geometry and axisymmetry in all our calculations. The 
velocity and magnetic fields can be written as
\begin{equation}
\mathbf{v}= \omega(r,\theta,t) r \sin\theta\mathbf{e}_{\phi} +  \mathbf{v}_m 
\end{equation}
\begin{equation}
\mathbf{B}=B(r,\theta,t) \mathbf{e}_{\phi} + \nabla \times \left[ 
   A(r,\theta,t) \mathbf{e}_{\phi}\right] ,
\end{equation}
where $B(r,\theta,t)$ and $A(r,\theta,t)$ represent the toroidal 
components of
the magnetic field and of the vector potential, respectively;
$\omega(r,\theta,t)\equiv\Omega-\Omega_0$ is the residual angular  velocity
and  $\mathbf{v}_m=v_r \mathbf{e}_r + v_{\theta} \mathbf{e}_{\theta}$  is the
meridional  circulation.

As a self-consistent calculation of the meridional circulation throughout the
dynamically coupled regions of the convective zone and the tachocline is beyond
the scope of this work, we will regard the meridional circulation as given. As
a result of the anelastic approximation, the circulation in the spherical 
shell may be represented by a stream function $\Psi$ so that the components of
the  meridional circulation can be written as
\begin{eqnarray}
v_r &=& \frac{1}{\rho r^2 \sin\theta} \,\partial_{\theta} \Psi \\
v_{\theta} &=& \frac{-1}{\rho r \sin\theta} \,\partial_r \Psi .
\label{eq:merid_circ}
\end{eqnarray}
In order to present more transparent equations we write the stream function in
the  following form:
\begin{equation}
\Psi = \psi(r) \sin^2\theta \cos\theta .
\label{eq:Psi}
\end{equation}
the form of the given function $\psi(r)$ specifying the flow.

The evolution of the poloidal magnetic field is, in contrast, self-consistently
calculated by integrating Eq.\,\ref{eq:induc} as usual:
\begin{equation}
\partial_t A + \frac{1}{r \sin\theta}(\mathbf{v}_m \cdot \nabla) 
  (r \sin\theta A) =  \eta \left( \nabla^2 - \frac{1}{r \sin\theta} \right) A.
\label{eq:pol}
\end{equation}

The components of the viscous stress tensor appearing in the azimuthal
component of the Navier-Stokes equation are
\begin{eqnarray}
  \tau_{\theta \phi} &=& \tau_{\phi \theta} = \rho \nu \frac{\sin\theta}{r}\, 
     \partial_\theta \left( \frac{v_{\phi}}{\sin\theta} \right) ,\\ 
  \tau_{\phi r} &=& \tau_{r \phi} =  \rho \nu r \,\partial_r \left( 
     \frac{v_{\phi}}{r} \right),
\end{eqnarray}
where $\nu$ is the (turbulent) viscosity. With this, the azimuthal components
of Eqs.\,\ref{eq:NS}--\ref{eq:induc}, including the effects of
diffusion, Coriolis force, meridional circulation, toroidal field production by
shear, and Lorentz force, read
\begin{eqnarray}
\label{eq:main1}
\partial_t \omega &=& \left( \partial_r \nu + 4 \frac{\nu}{r} + 
  \nu\,\frac{\partial_r\rho}{\rho} \right) \partial_r\omega
  + \frac{3 \nu \cot\theta}{ r^2} \,\partial_{\theta} \omega
   \\ \nonumber
  &+& \nu \,\partial^2_r \omega 
  +  \frac{\nu}{r^2}\, \partial^2_{\theta} \omega + L + M_1 + C  \\ \nonumber
 L &=& \frac{B}{4 \pi \rho r^2 \sin\theta} \left( \frac{\partial_{\theta} A}{r}
  - \cot\theta\,\partial_r A \right)  \\ \nonumber
  &+& \frac{\partial_r B}{4 \pi \rho r^2 \sin\theta} \left( 
   \partial_{\theta} A  - \cot\theta\,A \right)  \\ \nonumber
  &-& \frac{\partial_{\theta} B}{4\pi\rho r^2\sin\theta} 
  \left(\frac{A}{r}+\partial_r A \right)  \\ \nonumber
 M_1 &=& \frac{1}{r^2 \rho} \left( 2 \cos^2\theta\, \partial_r \psi + 
   \frac{1- 3\cos^2\theta}{r}\,2\psi \right)\,\omega \\ \nonumber
   &+& \dfrac{1- 3\cos^2\theta}{r^2 \rho}\,\psi\,\partial_r \omega 
   + \frac{\sin\theta\cos\theta}{r^2 \rho}  \,
   \partial_r \psi\,\partial_{\theta} \omega \\ \nonumber
 C &=& \frac{2\Omega_0}{r^3 \rho}\, ( 1-3 \cos^2\theta)\,\psi 
   + \frac{2\Omega_0\cos^2\theta }{r^2 \rho}\, \partial_r \psi  
\end{eqnarray}
\begin{eqnarray}
\label{eq:main2}
\partial_t B &=& \left( \frac{\partial_r \eta}{r} - 
  \frac{\eta}{r^2 \sin^2\theta} \right) B + \left( \frac{2 \eta}{r} + 
  \partial_r \eta \right) \partial_r B \\ \nonumber
  &+& \frac{\eta}{r^2}\,\cot\theta\,\partial_{\theta}B  +\eta\,\partial^2_r B
  +\frac{\eta}{r^2}\, \partial^2_{\theta} B + M_2  \\ \nonumber
 M_2 &=& \frac{\psi}{r^2 \rho} \left[(3\cos^2\theta-1)\left( \frac 1r 
     +\frac{\partial_r \rho}{\rho}\right)\right] B \\ \nonumber
     &-& \frac{\partial_r\psi}{r^2 \rho}\,\cos^2\theta\, B  
     +\frac{\psi}{r^2 \rho} (1-3\cos^2\theta)\,\partial_r B
\end{eqnarray}
while equation~(\ref{eq:pol}) can be written as
\begin{eqnarray}
\label{eq:main3}
\partial_t A &=& \frac{- \eta}{r^2 \sin^2\theta}\,A 
    + \frac{2 \eta}{r}\, \partial_r A 
    + \frac{\eta \cot\theta}{r^2}\, \partial_{\theta} A \\ \nonumber
    &+& \eta \, \partial^2_r A
    + \frac{\eta}{r^2} \, \partial^2_{\theta} A 
    + M_3 \\ \nonumber
  M_3 &=& \frac{1}{r^2 \rho} \left( \frac{1-3\cos^2\theta}{r}\psi + 
       \cos^2\theta\,\partial_r \psi \right)\,A \\ \nonumber
      &+& \frac{\psi}{r^2 \rho} (1-3\cos^2\theta)\,\partial_r A 
      + \frac{\sin\theta\cos\theta}{r^2 \rho}\,\partial_r \psi\,\
      \partial_{\theta} A 
\end{eqnarray}
In these formulae $L$ denotes the terms associated with the Lorentz force, $C$ denotes the
terms  associated with the Coriolis force and $M_1$, $M_2$, $M_3$ denotes the
terms associated  with the advection by meridional circulation.

\subsection{Boundary and initial conditions}

The computational domain for the present calculations consists of just the
upper part of the radiative interior, between radii $\rin$ and $\rbcz$, where
$\rbcz$  is the radius of the bottom of the convection zone. For the
integration of Eqs.\,\ref{eq:main1}--\ref{eq:main3}
we use the following boundary conditions.

At the pole and the equator axial and equatorial symmetry is required for
$\omega$, and dipole symmetry for the magnetic field:
\begin{eqnarray}
\partial_\theta \omega = B=\hspace{1 em}A=0  \hspace{2 cm} &&\mbox{at }
  \theta = 0, \nonumber \\ 
\partial_\theta \omega = B=\partial_\theta A=0  \hspace{2 cm} &&\mbox{at }
  \theta = \pi/2. 
\end{eqnarray}

The bottom of our box is supposed to be a rigidly rotating perfect conductor ---
a good approximation for the solar interior. The rigid rotation speed is fixed
at the observed rate $\Omega_0/2\pi = 437$ nHz. Thus, the lower boundary
conditions are:
\begin{equation}
\omega=B=A=0 \hspace{2 cm} \mbox{at }r = \rin
\end{equation}
Note that the fixed internal rotation rate $\Omega_0$ is in general not equal
to the rate at which the interior of our models would rotate if the lower
rotation rate were not prescribed. (In other words, the fixed rate does not in
general ensure zero net torque at the lower boundary.) This leads to a spurious
{\it radial} differential rotation in the radiative zone below the tachocline.
In models where this effect is strong, it is a sign that the model does not
represent well the observations as it would spontaneously lead to an internal
rotation rate that differs significantly from what is observed. As, however,
the magnetic field geometry considered here is still not realistic, firm
conclusions on this would be premature anyway.

The lower boundary needs also to be placed sufficiently deep in the solar
interior to make sure that the confinement of the tachocline is not an artefact
caused by the rigid rotation imposed at the bottom. For this reason, in most
calculations presented here we set $\rin=0.4 R_\odot$, even though only the
upper part of our computational domain is shown in many of the figures.

At the upper boundary of our domain, i.e. at the bottom of the convective zone,
we use the following boundary conditions:
\begin{eqnarray}
&&\omega=\Omegabcz-\Omega_0 \nonumber \\
&&A=A_0 \sin\theta \cos (\omegacyc t) \hspace{1 cm} \mbox{at } r = \rbcz, 
   \\ 
&&B=0 \nonumber
\end{eqnarray}
The physical assumptions behind these conditions are the following. For
$\omega$ we suppose that the rotation rate can be described with the same
expression as in the upper part of the convection zone. In accordance with the
observations of the GONG network, the expression used for
$\Omegabcz$ is
\begin{equation}
\frac{\Omegabcz}{2\pi} = 456 - 72 \cos^2\theta - 42 \cos^4\theta 
  \hspace{0.2cm} \mbox{nHz}.
\end{equation}
Our upper boundary condition on the poloidal field is the simplest possible
representation of the poloidal component of the solar dynamo field, a dipole
oscillating with a period of 22 years. The factor $A_0$ fixes the poloidal 
field amplitude. Finally, the upper boundary condition chosen for
the toroidal field assumes that the toroidal field is negligible in the
convective zone compared to its value in the tachocline. Such a hypothetical
situation is in line with current thinking in dynamo theory, and it may be a
natural consequence of buoyancy-driven instabilities effectively removing any
toroidal flux from the convective zone.

The initial conditions chosen for all calculations are
\begin{equation}
\begin{array}{lll}
\label{eq:incond}
\omega (r,\theta , t=0)  &=  \Omegabcz - \Omega_0 \hspace{0.6cm} &\mbox{at}
	\hspace{0.1cm}r = \rbcz \\
\omega (r,\theta , t=0)  &=  0  \hspace{1.3cm} &\mbox{at} 
	\hspace{0.1cm}r < \rbcz  \\
B (r,\theta , t=0)  &\equiv  0  &  \\ 
A (r,\theta , t=0)  &= A_0 \sin^2\theta  \hspace{0.64cm} &\mbox{at} 
	\hspace{0.1cm}r = \rbcz  \\ 
A (r,\theta , t=0)  &=  0  \hspace{1.8cm} &\mbox{at} 
	\hspace{0.1cm}r < \rbcz 
\end{array}
\end{equation}
Some test runs with other initial conditions (with significant differential
rotation rotation throughout the volume) were also performed. After stationary
periodic behaviour sets in, the results were found to be independent of the
initial conditions chosen.

   \begin{figure*}
   \centering
   \includegraphics[width=0.85\linewidth]{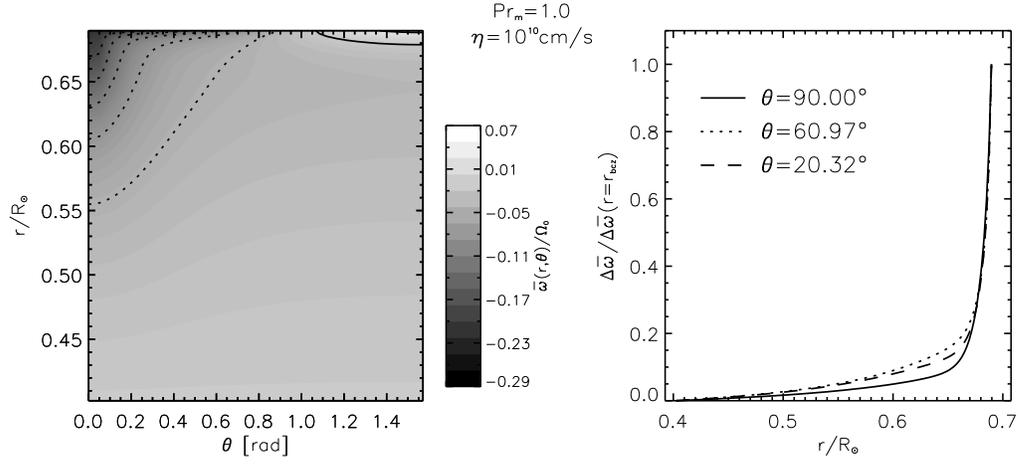}
      \caption{Spreading of the differential rotation into the radiative
	       interior for $\eta=\nu=10^{10} \mbox{cm}^2/\mbox{s}$.  {\it
	       Left-hand panel:} contours of the time-average of the angular
	       rotation rate  $\omega(r,\theta,t)$ under one dynamo period.
	       Equidistant contour levels are shown,  separated by intervals of
	       $100\,\mbox{nHz}/\Omega_0$, starting from 0 towards both 
	       non-negative (solid) and negative (dashed) values. {\it
	       Right-hand panel:} normalized differential  rotation amplitude
	       $\Delta\omega$ at different latitudes as a function of
	       radius.  The peak amplitude of the poloidal magnetic field 
	       is $B_p\sim2570$ G.
              }
         \label{fig:atlag_1a}
   \end{figure*}

   \begin{figure*}
   \centering
   \includegraphics[width=0.85\linewidth]{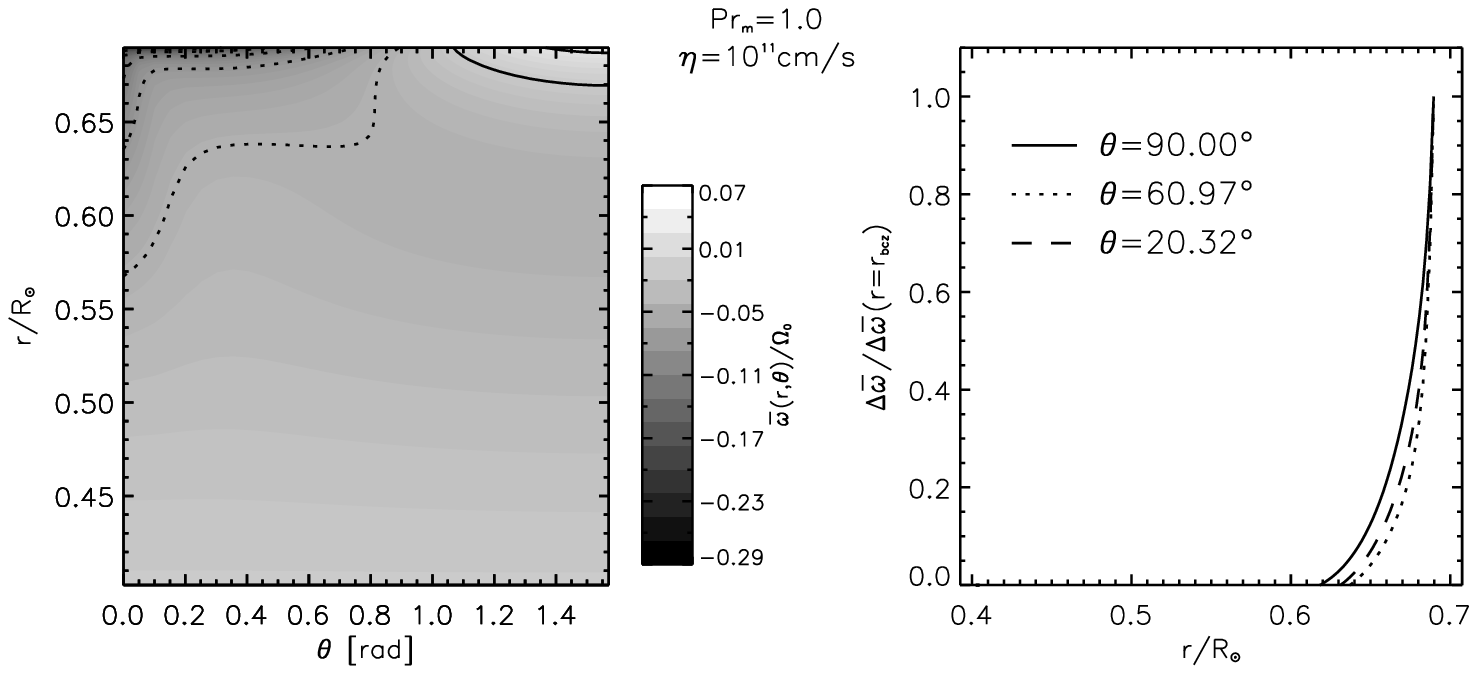}
      \caption{Same as in Fig.\,\ref{fig:atlag_1a} for $\eta=\nu=10^{11}
	       \mbox{cm}^2/\mbox{s}$.   The peak amplitude of the poloidal
	       magnetic field is $B_p\sim6660$ G.
              }
         \label{fig:atlag_2b}
   \end{figure*}

   \begin{figure*}
   \centering
   \includegraphics[width=0.85\linewidth]{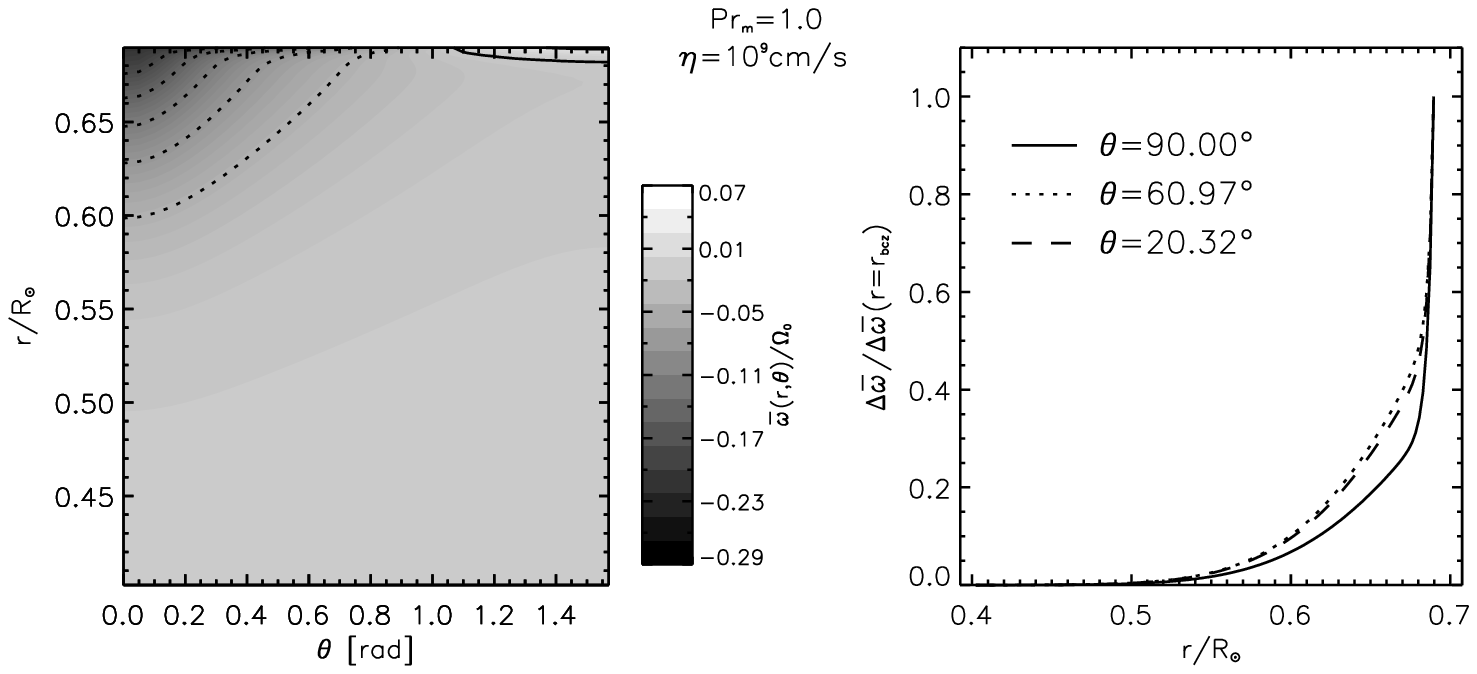}
      \caption{Same as in Fig.\,\ref{fig:atlag_1a} for $\eta=\nu=10^{9}
	       \mbox{cm}^2/\mbox{s}$.   The peak amplitude of the poloidal
	       magnetic field is $B_p\sim1386$ G.
              }
         \label{fig:atlag_3a}
   \end{figure*}

   \begin{figure*}[!hp]
   \centering
   \includegraphics[width=0.95\linewidth]{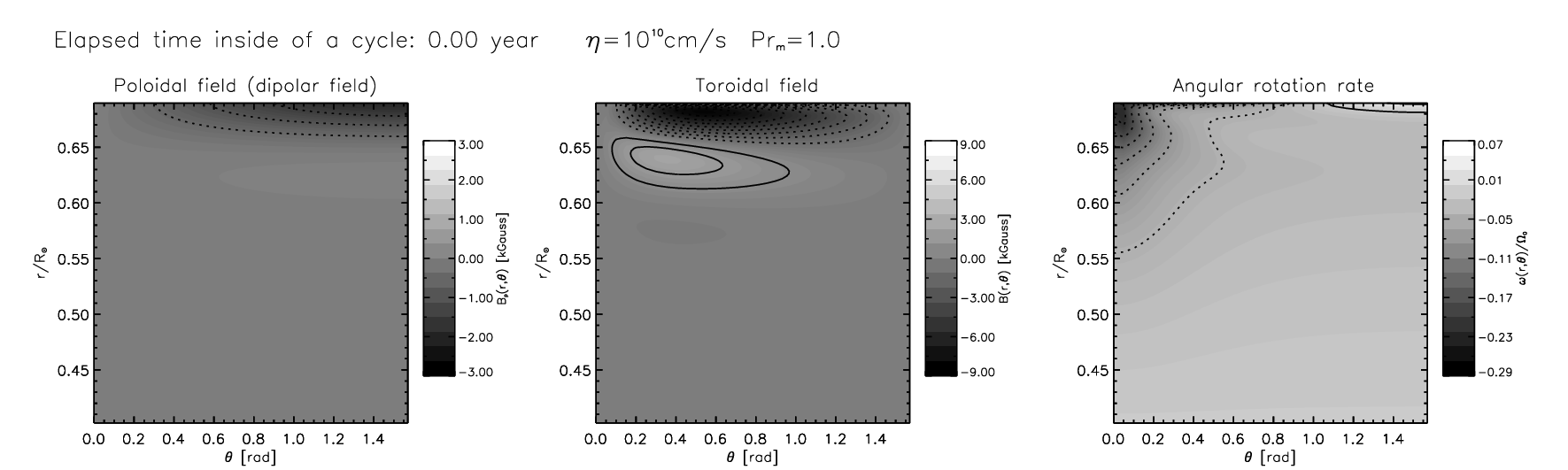}
   \includegraphics[width=0.95\linewidth]{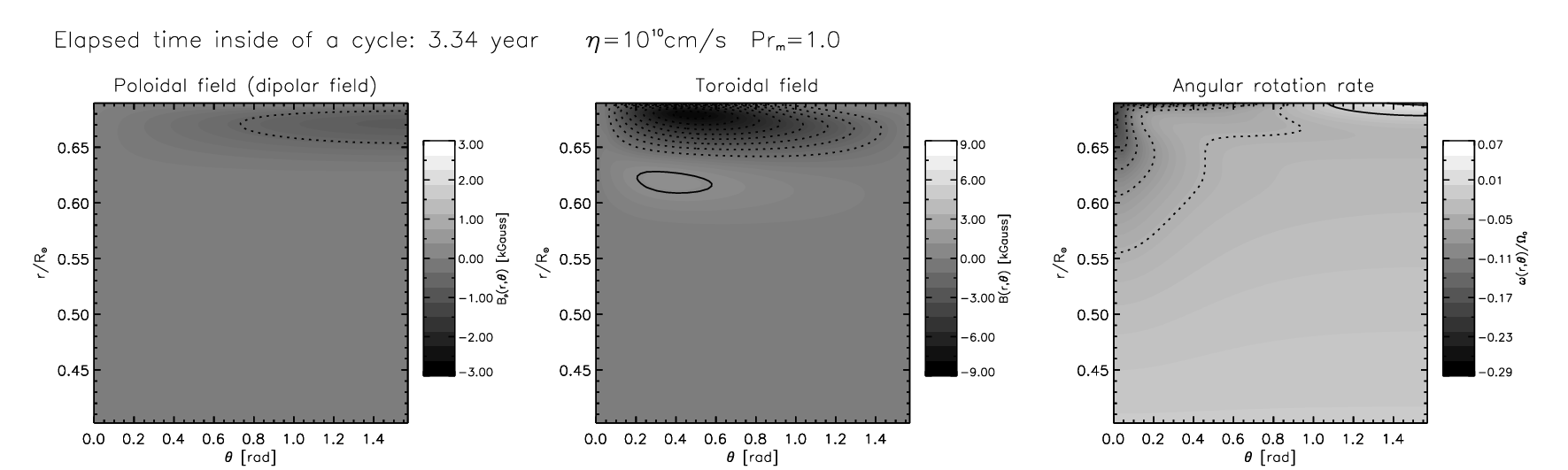}
   \includegraphics[width=0.95\linewidth]{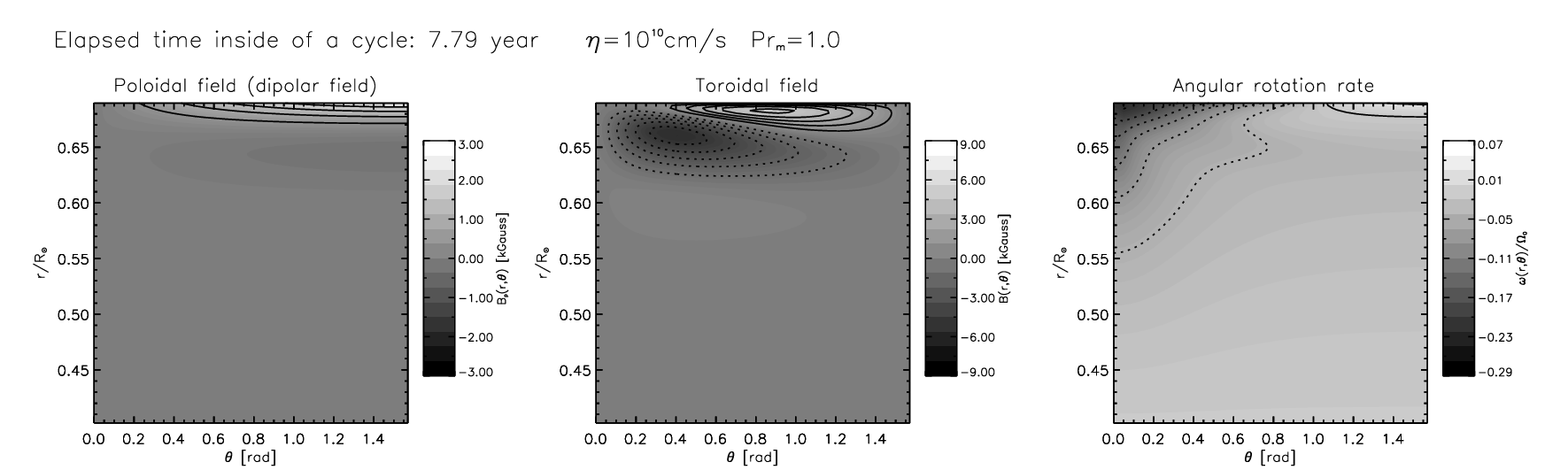}
   \includegraphics[width=0.95\linewidth]{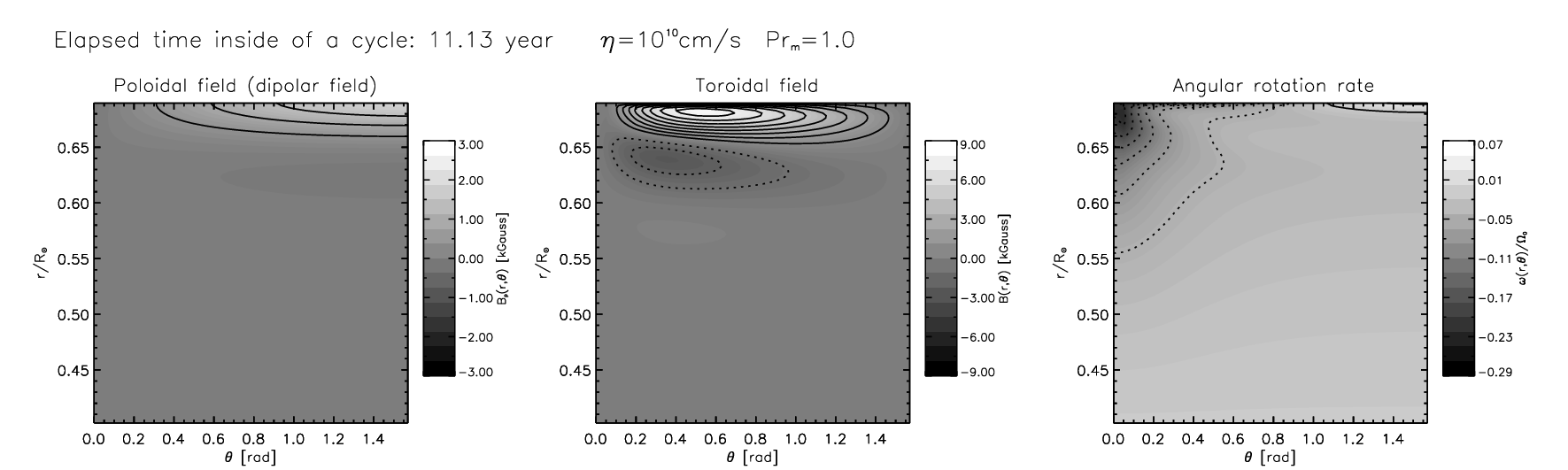}
      \caption{Snapshots of the solution at four cycle phases for
	       $\eta=\nu=10^{10} \mbox{cm}^2/\mbox{s}$.  {\it Left-hand panels:
	       } contours of poloidal magnetic field strength.   Equidistant
	       contour levels are shown, separated by intervals of $500\,$G.
	       {\it Middle panels: } contours of toroidal magnetic field
	       strength.  Equidistant contour levels are shown, separated by
	       intervals of $1000\,$G.  {\it Right-hand panels: } contours of
	       angular rotation rate, as in Fig.\,\ref{fig:atlag_1a}.
              }
         \label{fig:fejlodes}
   \end{figure*}

\subsection{Numerical method}

We used a time relaxation method with a finite difference scheme first order
accurate in time to solve the equations. A uniformly spaced grid is set up
with 128 grid points in the $r$ direction and 32 grid points in the $\theta$ direction.

Our calculations are based on a more recent version of the solar model of 
\citet{Guenther}.

Starting from the initial conditions (\ref{eq:incond}), the solution is
allowed to evolve in time until it relaxes to a very nearly periodic
behaviour. The timescale needed for this is the diffusive timescale
\begin{equation}
t_{\mbox{\scriptsize dif}} = \frac{(\rbcz-\rin)^2}{\mbox{Min}(\nu,\eta)} ,
\end{equation}
where $\mbox{Min}(\nu,\eta)$ is the lowest value of viscosity or magnetic
diffusivity in the domain.  Physically we would expect this value to be close
to the molecular viscosity. Using this value in the computations would,
however, lead to a prohibitively high number of timesteps to relaxation.
(Clearly, the runtime of  our computations must be chosen to well exceed 
$t_{\mbox{\scriptsize dif}}$ to reach relaxation.) Consequently, in the
present paper we only consider cases where the diffusivities are constant
throughout the domain. Note, however, that in FDP01 a case with the diffusivity
significantly reduced towards the interior was also computed, and no
qualitative changes in the character of the solution were found. 

The question may arise, to what extent does the use of such a high
viscosity in the deep radiative zone distort the results? The answer is clearly
that the main difference between models with high and low diffusivities is the
diffusive timescale over which the model evolves to equilibrium. Once
equilibrium {\it is} reached, the azimuthal component of the equation of motion
reduces to a Laplace equation (neglecting meridional circulation and magnetic
fields), so the solution is universal and independent of the diffusivity. The
equilibrium solution is thus not expected to be strongly influenced.

On the other hand, for realistically low (such as molecular) diffusivities
equilibration takes much longer than the age of the Sun. Thus, an internal
differential rotation, induced by solar wind torques or as a pre-MS relict, may
not be suppressed. It is therefore important to make it clear (as we already
did in FDP01) that the present work does {\it not} propose that
dynamo-generated fields can explain the overall lack of differential rotation
in the deep radiative zone, just that they dominate the dynamics of the
tachocline itself. The lack of differential rotation in the deep radiative zone
still needs to be explained by an internal magnetic field independent of the
dynamo.

\section{Results and discussion}

\subsection{Reference case} 

First we present results for a case with $\Prm=1$,  $\eta=\nu=10^{10}
\mbox{cm}^2/\mbox{s}$, and no meridional circulation ($\psi=0$). This
``reference case'' corresponds to that computed in FDP01, except that the
poloidal field is now not prescribed throughout the volume, but explicitly
calculated. 
The results are shown in Fig.\,\ref{fig:atlag_1a}. 

In the left-hand panels we plotted the contours of the time-average of 
the angular rotation rate, which is defined as
\begin{equation}
\overline{\omega}(r,\theta) = \frac 1{\Pcyc}\int_t^{t+\Pcyc}{\omega(r,\theta,t)}\,\mathrm{d}t.
\end{equation}

\clearpage
In the right-hand panels the differential rotation amplitude 
$\Delta \omega$ is defined as
\begin{equation}
\Delta \omega = \omega(r,\theta,t) - \int_0^{\pi/2}\omega(r,\theta,t) 
   \sin \theta \,\mathrm{d}\theta 
\end{equation}
\begin{equation}
\Delta \overline{\omega} = \overline{\omega}(r,\theta) - \int_0^{\pi/2} \overline{\omega}(r,\theta) 
   \sin \theta \,\mathrm{d}\theta . 
\label{eq:delta_omega}
\end{equation}
The thickness of the tachocline will be defined throughout this paper as the 
scale height of $\Delta\omega$ (i.e.\  the depth from the top of our domain
where $\Delta\omega$ is reduced by 1/e). By ``mean thickness'', in turn, we
will refer to the scale height of $\Delta\ov\omega$.

The results shown in Fig.\,\ref{fig:atlag_1a} are in accordance with those
in FDP01. In particular, we confirm that an oscillatory poloidal field with
dipolar latitude dependence at the tachocline--convective zone boundary is able
to confine the thickness of tachocline. The effectivity of this confinement
increases (i.e.\ the thickness of the tachocline decreases) with the field
strength. 

This conclusion is basically valid for all the cases studied in this paper,
though the field strength necessary to just reproduce the observed  
equatorial mean thickness of the tachocline, to be called the {\it confining
field strength} $\Bconf$, for brevity, varies from case to case, as do the
structural details of the resulting tachocline. Though in order to find the
right value of $\Bconf$ we experimented with many different field strengths for
each parameter combination, in this paper we only present results computed with
$B=\Bconf$ (cf. FDP01 for calculations with other field strengths).
The value of the confining field strength is found to be $\sim 2600$ G 
for the reference case, very close to the value found in FDP01. 

In what follows, we will systematically study the effects of varying the
parameters of the model on the resulting tachocline structure and on the value
of the confining 
field strength. First, we fix the value of the magnetic Prandtl number as 
$\Prm=\nu/\eta=1.0$, while   we change the amplitude of the magnetic
diffusivity. Second, we examine the influence of varying the magnetic Prandtl
number on the radial spreading of the differential rotation into the radiative
interior. In these two  cases we ignore the meridional flow.  Finally, we will
focus on the effect of introducing a  meridional flow. 

Computer animations illustrating the time development of some of our solutions
can be downloaded from the following web site:\newline
{\tt http://astro.elte.hu/kutat/sol/fast1/fast1e.html}

   \begin{figure}
   \centering
   \includegraphics[width=0.95\linewidth]{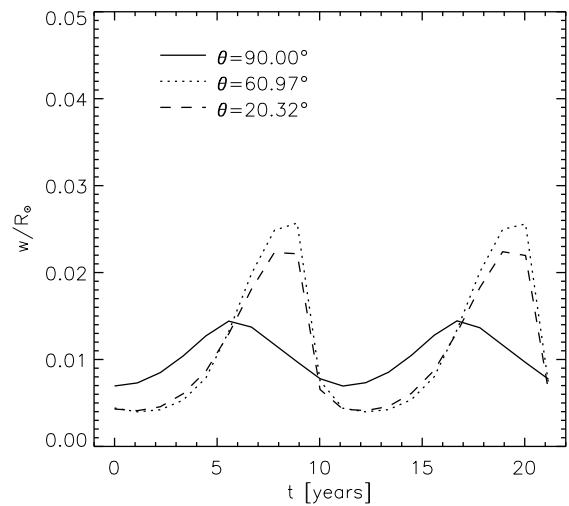}
      \caption{The thickness of the tachocline at different latitudes as a 
               function of time
	       for the case in  Fig.\,\ref{fig:atlag_1a}.
	       }
         \label{fig:w_ido_1a}
   \end{figure}

   \begin{figure}[!t]
   \centering
   \includegraphics[width=0.95\linewidth]{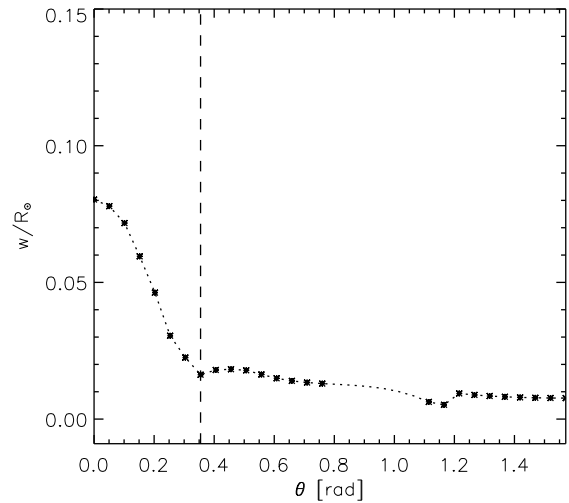}
      \caption{Latitudinal variation of the mean tachocline thickness for the
	       case in  Fig.\,\ref{fig:atlag_1a}. A vertical dashed line
	       indicates the poleward limit of reliable helioseismic data.
	       }
         \label{fig:w_th_1a}
   \end{figure}

\subsection{Varying the diffusivity}

In this subsection, the value of the magnetic Prandtl number was fixed,  
while we studied the effects of varying the diffusivity. 
The results are shown in Figs.\,\ref{fig:atlag_1a}-\ref{fig:atlag_3a}.

   \begin{figure*}
   \centering
   \includegraphics[width=0.85\linewidth]{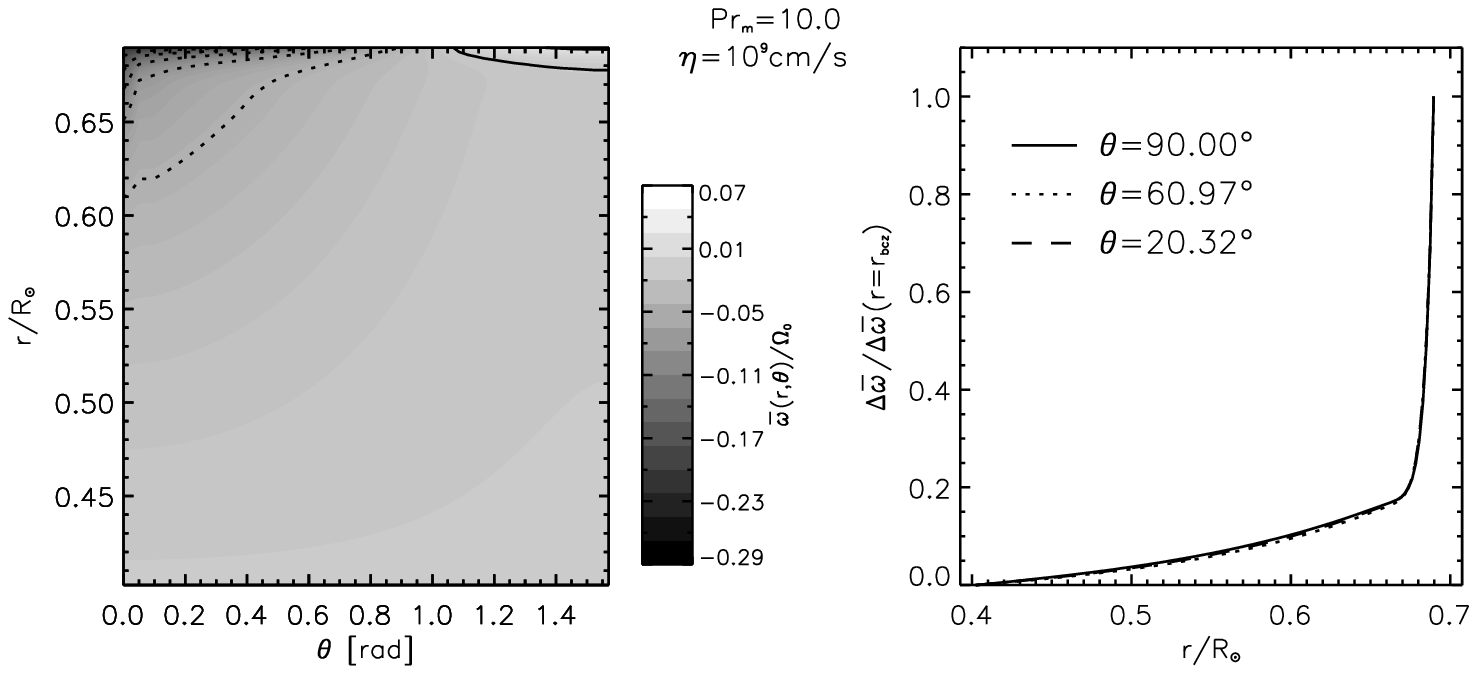}
      \caption{Same as in Fig.\,\ref{fig:atlag_1a} for $\Prm=10.0$, where
	       $\eta=10^{9} \mbox{cm}^2/\mbox{s}$  and $\nu=10^{10}
	       \mbox{cm}^2/\mbox{s}$.   The peak amplitude of the poloidal
	       magnetic field is $B_p\simeq 8531$ G.
              }
         \label{fig:atlag_4a}
   \end{figure*}

   \begin{figure*}
   \centering
   \includegraphics[width=0.85\linewidth]{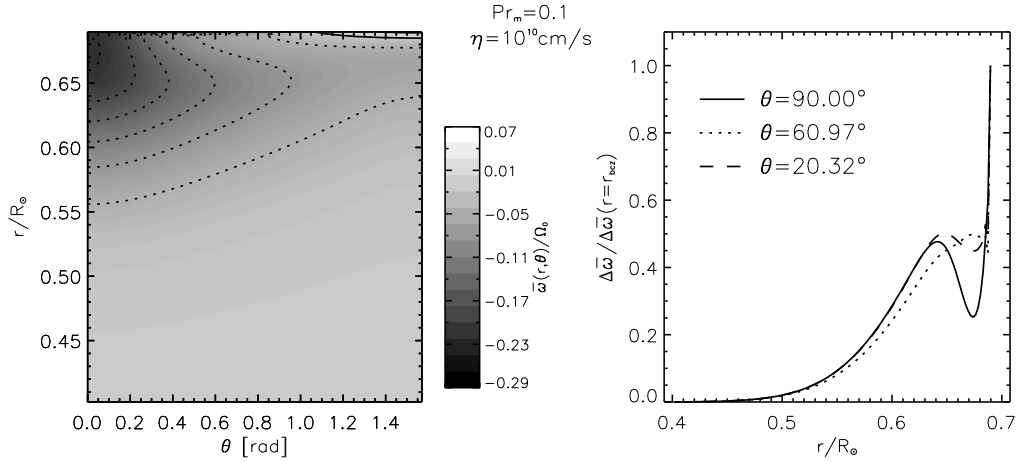}
      \caption{Same as in Fig.\,\ref{fig:atlag_1a} for $\Prm=0.1$, where
	       $\eta=10^{10} \mbox{cm}^2/\mbox{s}$  and $\nu=10^{9}
	       \mbox{cm}^2/\mbox{s}$.   The peak amplitude of the poloidal
	       magnetic field is $B_p\simeq 1028$ G.
              }
         \label{fig:atlag_5a}
   \end{figure*}

In line with the analytic estimates of FDP01, we find that $\Bconf$ increases
with the value of the diffusivity. Given that the equipartition field strength
in the deep convective zone is $\sim 10^4\,$G, the value of $\Bconf$ is quite
reasonable even with a diffusivity of $10^{11}\,$cm$^2/$s.

Beside increasing the value of $\Bconf$, a higher diffusivity has other effects
on the solution, too. The fast tachocline shows a quite marked time- and
latitude-dependence in our models, as illustrated in Fig.\,\ref{fig:fejlodes} 
for the reference case. As a consequence, the thickness of the tachocline
depends both on latitude and cycle phase. In particular, the tachocline tends
to be significantly wider at higher latitudes on the mean, while it is also
more time-dependent near the poles (Fig.\,\ref{fig:w_ido_1a}). While the
relative shortness of seismic data sets does not permit firm conclusions about
the cycle dependence of tachocline properties yet, the overall form of the mean
thickness--latitude relation (Fig.\,\ref{fig:w_th_1a}) is in good qualitative
agreement with the available helioseismic information \citep[cf. Fig.\,9
in][]{Basu+Antia:tachovar}. The amplitude of these variations within the
latitude interval where the helioseismic results are reliable ($\theta\ga
20^\circ$) also seems to be in agreement with the observations. 

For a fixed value of the field strength, a higher value of diffusivity tends to
reduce the amplitude of the spatiotemporal variations. On the other hand, for
higher diffusivities $\Bconf$ is also higher, so for this field strength a
stronger dynamical variation is imposed on the flow. Nevertheless, at least
for the reference case, the time dependence is well within the observational 
errors as shown e.g. in Fig.\,4 of \cite{Basu+Antia:tachovar}.

It must be noted of course that, as our imposed upper boundary condition on the
poloidal field, a simple oscillating dipole, may not be very realistic, the
detailed spatio-temporal properties of the solution should be regarded with
caution. Among such features we may mention the poleward drift of the resulting
rotational pattern, apparent e.g.\ in the phase lag between different latitudes
in Fig.\,\ref{fig:w_ido_1a}. (Cf.~the same trend in Fig.\,10 of FDP01.)

\subsection{Varying the magnetic Prandtl number}

Figs.\,\ref{fig:atlag_4a} and \ref{fig:atlag_5a} illustrate the effect of
magnetic Prandtl numbers other than 1 on the solution. As expected on the basis
of the analytical estimates of FDP01, tachocline confinement is less efficient
for high  magnetic Prandtl numbers, so that higher values of $\Bconf$ are
necessary if $\Prm>1$. On the other hand, a higher magnetic Prandtl number
(i.e. a relatively higher viscosity) is in general beneficial for the overall
smoothness of the solution, leading to less strong dependence of $\omega$ on
time and latitude. In the low $\Prm$ case, in turn, strong spatiotemporal
variations in $\omega$ become prevalent, leading to a complex non-monotonic
behaviour of the rotational velocity as a function of radius and latitude.

   \begin{figure}
   \centering
   \includegraphics[width=0.95\linewidth]{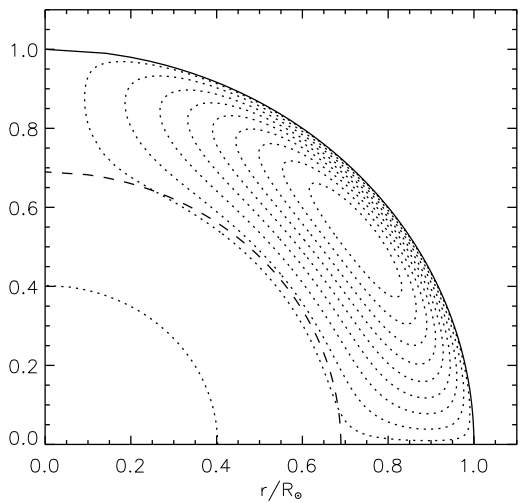}
      \caption{Streamlines of the meridional circulation prescribed in Eq.\,
	       (\ref{eq:psi_r}),  with the following parameter values:
	       $\psi_0=-10^{20}$, $k=1$,   $r_0=(R_{\odot}-\rmc)/30$ and
	       $\Gamma=3.4 \times 10^{10}$. Dotted lines represent 
	       counterclockwise circulation. The amplitude of the flow 
	       is $\sim 3$ cm/s in our computational domain (between the
	       dashed and dotted circles).
              }
         \label{fig:stream_counterclock}
   \end{figure}

   \begin{figure}
   \centering
   \includegraphics[width=0.95\linewidth]{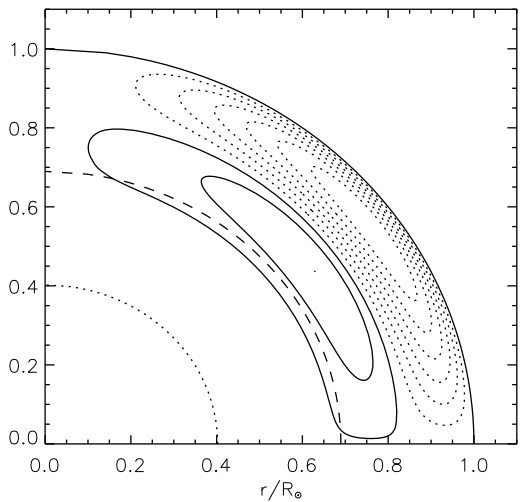}
      \caption{Same as in Fig.\,\ref{fig:stream_counterclock} with $\psi_0=10^{20}$,
	       $k=2$. Solid and dotted lines represent clockwise and
	       counterclockwise circulation, respectively.
	       }
         \label{fig:stream_clock}
   \end{figure}

\subsection{Influence of the meridional flow}

In the case of a fast tachocline dynamically coupled to the convective zone,
as assumed here, we expect that any meridional circulation present in the
tachocline is unseparable from the complex problem of meridional circulation 
in the convective zone. It is far beyond the scope of this paper to attempt to
model this circulation in a consistent way. Instead, we will limit ourselves to
regard the effects of a meridional circulation with a simple and plausible,
prescribed spatial structure, and a realistic amplitude.

It is known from direct and seismic studies that the meridional circulation
near the photosphere is poleward, with a peak amplitude of $\sim 20\,$m/s 
\citep{Komm+:circul,Latushko:circul}. If we assume that the
meridional flow in the convective zone$+$tachocline system follows a
simple one-cell pattern, the circulation in the tachocline will be equatorward.
The alternative possibility of a two-cell pattern with poleward flow in the
tachocline, however, should also be left open for now \citep{Kuker+Stix:circul}. 
An expression for the function $\psi(r)$, defined in Eq.\,\ref{eq:Psi}, 
that encompasses both possibilities is 
\begin{equation}
\psi(r) = \psi_0 \sin \left[ \frac{k \pi (r-\rmc)}{R_{\odot}-\rmc} \right] 
          \exp \left(\frac{(r-r_0)^2}{\Gamma^2} \right)  ,
\label{eq:psi_r}
\end{equation}
where $\psi_0$ sets the amplitude of the meridional circulation, $r_0$, 
$\Gamma$ and $k$ are geometric parameters and $\rmc$ is the radius value 
down to which the meridional flow penetrates from the base of the convective
zone. Keeping in mind the observed depth of tachocline, here we set
$\rmc=4.6\times 10^{10}\,$cm.

The strong subadiabatic stratification of the solar interior sets an upper
limit to the amplitude of the meridional flow. The timescale of any meridional
circulation cannot be shorter than the relevant thermal diffusive timescale (to
allow moving fluid elements to get rid of their buoyancy). In the present case
the relevant heat conductivity is the turbulent one, so for turbulent Prandtl
numbers not very different from unity, the shortest possible timescale for
meridional circulation is just comparable to the viscous timescale. An
elementary estimate yields $v_{\mbox{\scri max}}\sim\rbcz/t_{\mbox{\scriptsize
dif}} \sim 10\,$cm$/$s.

The flow parameters used in our calculations were chosen to produce simple
smooth one- and two-celled flow patterns obeying the amplitude constraint
discussed above, while reproducing the observed flow speed near the surface.
The flow patterns are shown in Figs.\,\ref{fig:stream_counterclock} and
\ref{fig:stream_clock}. In these cases, the speed of the horizontal motion is $\sim
3\,$cm/s in the upper part of the radiative interior.

\begin{figure*}
   \centering
   \includegraphics[width=0.85\linewidth]{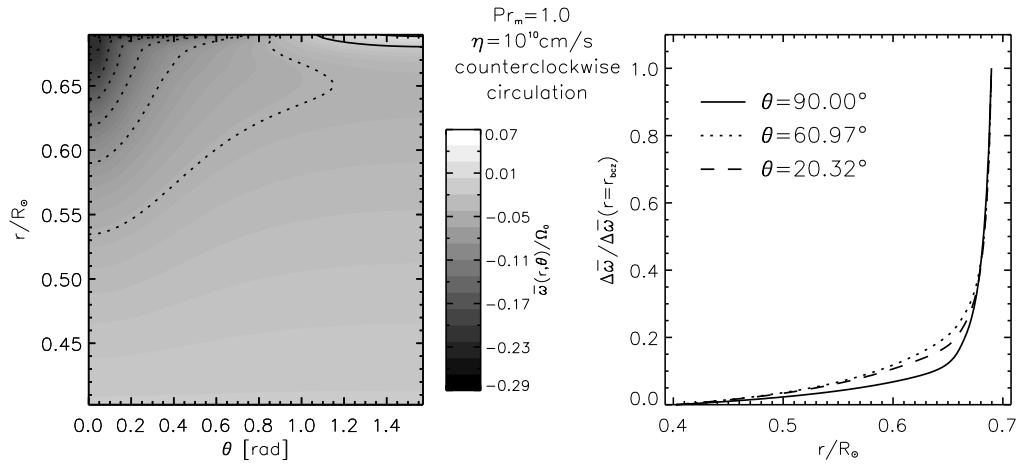}
      \caption{Same as in Fig.\,\ref{fig:atlag_1a}, but including the
	       meridional  circulation shown in Fig.\,\ref{fig:stream_counterclock}.
	      }
         \label{fig:atlag_counterclock_a}
   \end{figure*}

\begin{figure*}
   \centering
   \includegraphics[width=0.85\linewidth]{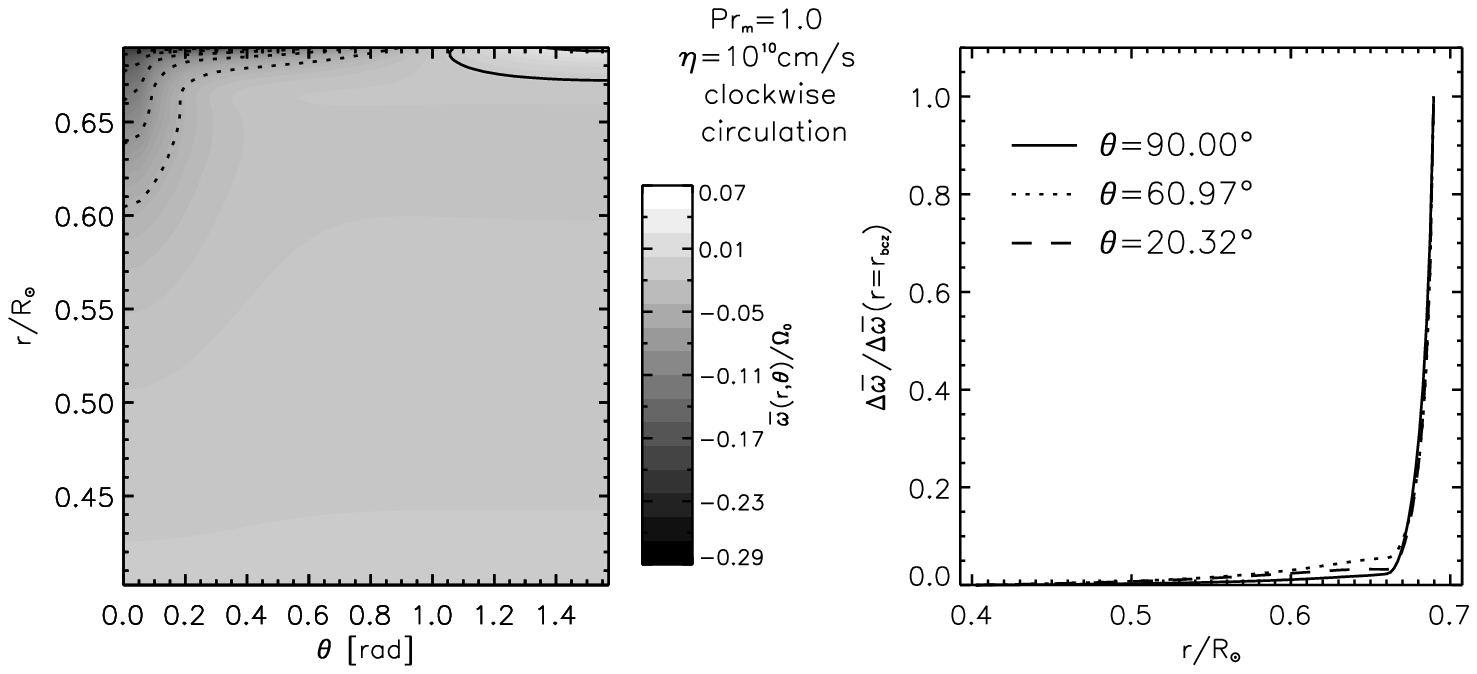}
      \caption{Same as in Fig.\,\ref{fig:atlag_1a}, but including the
	       meridional  circulation shown in Fig.\,\ref{fig:stream_clock}.
              }
         \label{fig:atlag_clock_a}
   \end{figure*}

The results with the meridional circulation included can be seen in
Figs.\,\ref{fig:atlag_counterclock_a} and \ref{fig:atlag_clock_a}. The
poloidal field amplitude used here was the same as in the reference case. It is
apparent that an equatorward flow in the tachocline region reduces somewhat the
efficiency of the confinement (the tachocline becomes slightly thicker), though
the effect is not dramatic. A poleward flow, on the other hand, significantly
improves the efficiency of the confinement. 

\begin{figure*}
   \centering
   \includegraphics[width=0.85\linewidth]{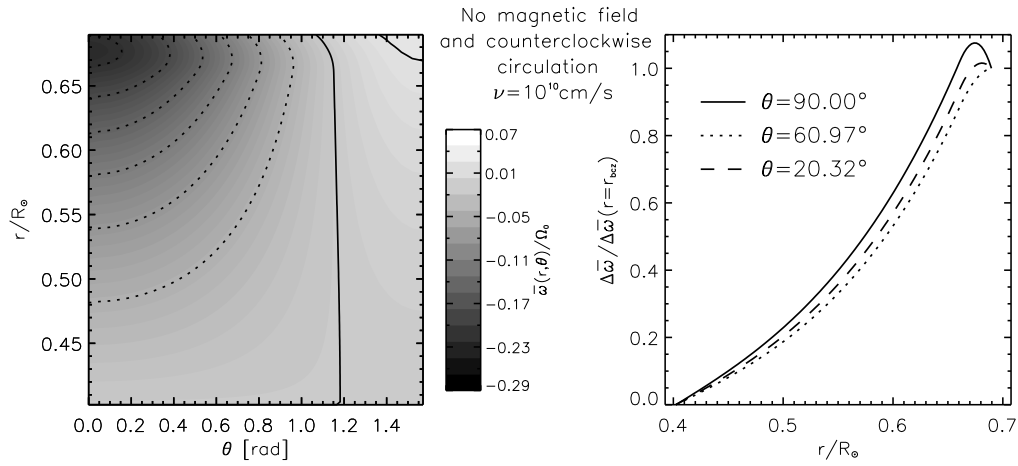}
      \caption{Spreading of the differential rotation into the radiative
	       interior including the
	       meridional  circulation shown in Fig.\,\ref{fig:stream_counterclock}, but 
	       taking the imposed field to be null.
	      }
         \label{fig:atlag_counterclock_nomag}
   \end{figure*}

\begin{figure*}
   \centering
   \includegraphics[width=0.85\linewidth]{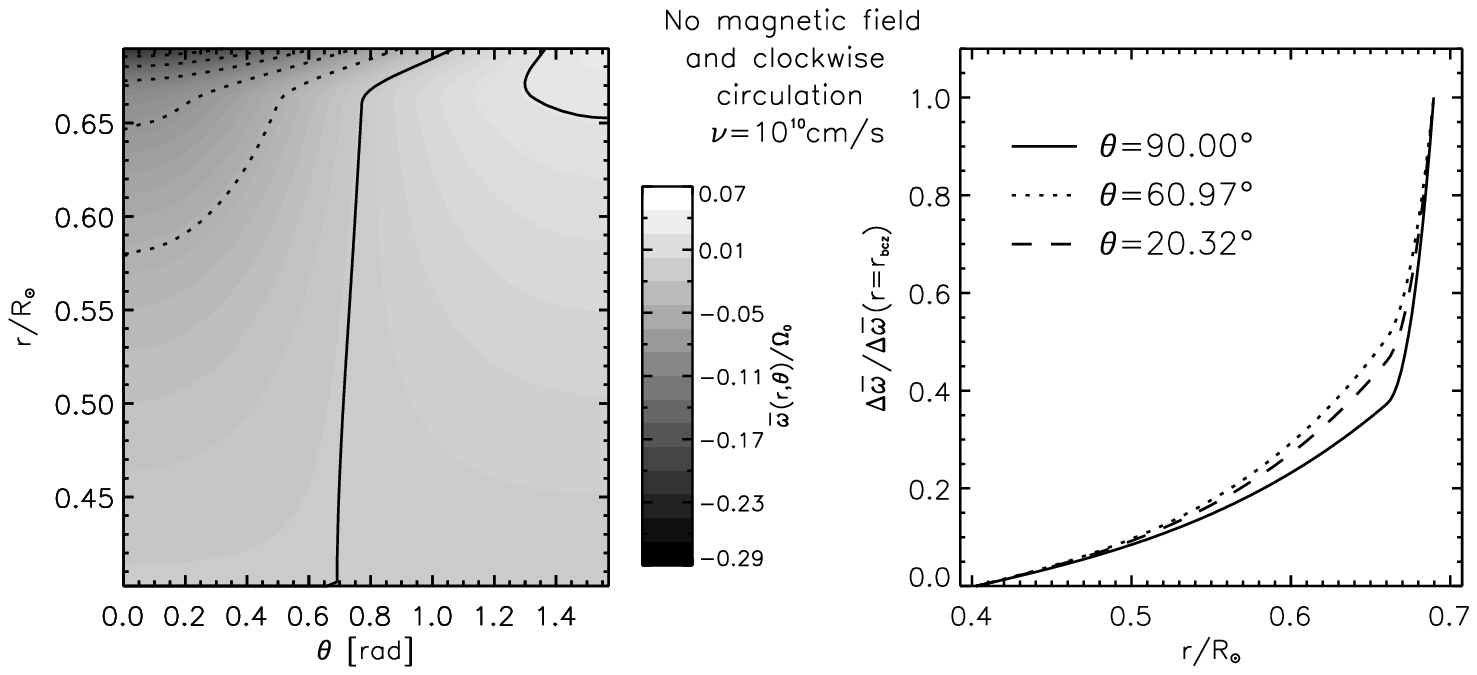}
      \caption{Same as in Fig.\,\ref{fig:atlag_counterclock_nomag} for the
	       meridional  circulation shown in Fig.\,\ref{fig:stream_clock}.
              }
         \label{fig:atlag_clock_nomag}
   \end{figure*}

In fact, meridional circulation alone, without any imposed magnetic field, can
also have a significant effect on the thickness of the tachocline, as it can
be seen by comparing Figs.\,\ref{fig:atlag_counterclock_nomag} and 
\ref{fig:atlag_clock_nomag}.

The relative importance of meridional circulation scales approximately with the
ratio of the  diffusive/advective timescales $v/\nu$; thus, for a higher value
of $\nu$, a proportionally higher flow speed is needed to reproduce the same
tachocline thickness.

\section{Conclusion}

We have begun a systematic study of the dynamics of the fast solar tachocline
arising under the assumption that the turbulent diffusivity in the tachocline
exceeds $\eta\ga 10^9\,$cm$^2/$s. We extended the earlier preliminary analysis
of FDP01 by consistently computing the evolution of the poloidal field, and by
exploring the three-dimensional parameter space defined by the viscosity in the
range $\log\nu=9$--$11$, the magnetic Prandtl number in the range
$\Prm=0.1$--$10$, and the meridional flow amplitude ($-3$ to $+3\,$cm$/$s).
The feedback of the Lorentz force to the meridional circulation was
neglected.

Confirming the results in FDP01, we found that, for basically all parameter
combinations considered, an oscillatory poloidal field with dipolar latitude
dependence at the tachocline--convective zone boundary is able to confine the
thickness of tachocline. (Note, however, that for low magnetic Prandtl numbers
the flow becomes rather complex, and a simple, smooth, monotonic tachocline
layer does not form.) The effectivity of this confinement increases (i.e.\ the
thickness of the tachocline decreases) with the field strength. The {\it
confining field strength} $\Bconf$ necessary to just reproduce the observed 
mean equatorial thickness of the tachocline is found to increase with
increasing viscosity, magnetic Prandtl number $\nu/\eta$, and meridional flow
speed (if considered positive equatorwards). Nevertheless, the resulting
$\Bconf$ values remain quite reasonable, in the range $10^3$--$10^4\,$G, for
all parameter combinations considered here.

The detailed dynamics of the solutions indicates that the thickness of the
tachocline, in general, increases with latitude, and it also shows a rather strong
dependence on cycle phase (Figs.\,\ref{fig:w_ido_1a} and \ref{fig:w_th_1a}). 
The latitude dependence is similar to that inferred by
helioseismology, while the time dependence is within the observational errors.

The models with no meridional circulation also show a tendency for the
rotational modulation to drift polewards (cf.\ Fig.\,\ref{fig:w_ido_1a}). It
must be noted, however, that owing to the rather simplified geometry of the
poloidal magnetic field imposed on the upper boundary (oscillating dipole), the
relevance of our results to the real Sun is dubious, as far as the detailed
spatiotemporal structure of the tachocline is concerned. Another example of
such features is the polar ``pit'' of slow rotation in nearly all
of our models: again, the reality of this pit needs to be verified with
other poloidal field geometries. A next step in our study of the fast
tachocline should clearly be an extension of this investigation to more general
magnetic field structures, reminiscent of the butterfly diagram.

Another important shortfall of these models is obviously the extremely
simplified treatment of the turbulent transport, by constant scalar
diffusivities. The turbulence present in the tachocline is presumably the
product of complex MHD instabilities. The study of the nonlinear development of
these instabilities has just begun \citep{Cally,Miesch:tacho1}, and we are
clearly a long way from being able to give a consistent description of
tachocline turbulence. Nevertheless, the present models could still be
significantly improved by considering simple, generic cases such as a scalar
diffusivity coupled to the value of the local shear. Extensions of the present
work along these lines will be treated in upcoming papers of this series.

\acknowledgements 
The authors wish to thank the referee, Pascale Garaud, for helpful suggestions.
We are also grateful to Sarbani Basu for useful insights on how to interpret
helioseismic constraints on the tachocline, and to D. B. Guenther for making
his solar  model available. This work was supported in part by the National
Science Foundation under grant no.\ PHY99-07949 and by the OTKA under grant 
no.\ T032462.

%
%______________________________________________________________

\end{document}